\documentclass[a4paper,11pt,oneside]{article}

\usepackage[english]{babel}
\usepackage[T1]{fontenc}
\usepackage[utf8]{inputenc}

\usepackage[pdftex,unicode]{hyperref}
\hypersetup{pdftitle=Product Relation Correlation and Its Use in Product Clustering}
\hypersetup{pdfauthor=Ondřej Sokol and Vladimír Holý}

\usepackage{color}
\definecolor{mycolor}{rgb}{0.267,0.004,0.329}
\hypersetup{colorlinks=true, linkcolor=mycolor, anchorcolor=mycolor, citecolor=mycolor, filecolor=mycolor, urlcolor=mycolor}
\definecolor{light-gray}{gray}{0.65}

\usepackage[margin=60pt]{geometry}
\usepackage[authoryear]{natbib}

\usepackage{amsmath}
\usepackage{amssymb}
\usepackage{graphicx}
\usepackage[group-minimum-digits=3]{siunitx}

\begin{document}

\begin{center}
{\Large \bfseries Product Relation Correlation and Its Use in Product Clustering}
\end{center}

\begin{center}
{\bfseries Petr Krautwurm} \\
Prague University of Economics and Business \\
Winston Churchill Square 4, 130 67 Prague 3, Czechia \\
\href{mailto:petr.krautwurm@vse.cz}{petr.krautwurm@vse.cz} \\
\end{center}

\begin{center}
{\bfseries Ondřej Sokol} \\
Prague University of Economics and Business \\
Winston Churchill Square 4, 130 67 Prague 3, Czechia \\
\href{mailto:ondrej.sokol@vse.cz}{ondrej.sokol@vse.cz} \\
\end{center}

\begin{center}
{\bfseries Vladimír Holý} \\
Prague University of Economics and Business \\
Winston Churchill Square 4, 130 67 Prague 3, Czechia \\
\href{mailto:vladimir.holy@vse.cz}{vladimir.holy@vse.cz} \\
\end{center}

\noindent
\textbf{Abstract:}
This paper introduces product relation correlation, a measure of product relatedness that assesses the extent to which products may function as substitutes or complements through analysis of shared purchasing patterns. Product relation correlation can be used for tasks such as product clustering and shelf space optimization, enabling retailers to arrange items in ways that enhance customer experience. Applied to data from a  retail drugstore chain, the measure demonstrates an alignment with cross-price elasticity, increasing as products diverge from independence. With computational simplicity, requirement for only commonly available data, and a robust theoretical interpretation, product relation correlation serves as a practical and efficient tool for deriving useful product insights.
\\

\noindent
\textbf{Keywords:} Retail Business, Scanner Data, Market Structure Analysis, Substitutes, Complements, Hierarchical Clustering.
\\

\noindent
\textbf{JEL Codes:} C38, D12, M31.
\\

\section{Introduction}
\label{sec:intro}

In the retail industry, especially in brick-and-mortar stores, customer satisfaction is significantly influenced by the ability to select and purchase appropriate products (see \citealp{Andreasen1977}). Achieving this requires that customers have access to comprehensive information for assessing product compatibility with their shopping objectives. This information can be acquired either in advance, as indicated by studies on the informational value of advertising (see, e.g., \citealp{Nelson1970, Nelson1974}), or during the shopping process, as suggested by research on the impact of shelf management in retail layouts (see, e.g., \citealp{Dreze1994, Grandi2021}). In this context, customers with prior knowledge typically enter the market with a specific purchase in mind, making them less responsive to in-store information, unlike those without predetermined purchase plan (see, e.g., \citealp{Lee2006, Grandi2021}). Moreover, customers without a predetermined purchase plan must make decisions on-site, which can be inherently more challenging due to constraints such as limited parking time or crowded shopping spaces. As a result, the satisfaction of customers who make in-store decisions is heavily dependent on the retailer’s ability to present suitable products efficiently, for instance, through the strategic clustering of related products, enabling customers to quickly compare and evaluate the best product options for their needs.

At its core, we define the relation between products as the extent to which they interact with one another. These interactions capture how much one product is required for the use or satisfaction of another. In essence, product relations can be understood as whether items are used together or are interchangeable.  Accordingly, products can be categorized as substitutes, complements, or independent (for a more detailed description, see, e.g., \citealp{Kort2020}). Strong relation between products imply that products are either substitutes or complements, while weak relations suggest they are closer to being independent. This interpretation is broadly consistent across multiple research domains (see, e.g., \citealp{Bakos1999, Manchanda1999, Shocker2004, Leeflang2012, Netzer2012, Kim2012})\footnote{Beyond retail, the closeness of products can also be utilized in finance through the pairs trading strategy, in which pairs of related stocks, such as Coca-Cola and Pepsi, are traded (see \citealp{Holy2025e})}.

Although customers often consider the specific nature of product relations in their decision-making, we propose that for retailers, the mere knowledge of a relation between products may be, in certain instances, sufficient for effective clustering. This is because highly related products elicit two contrasting behavioral responses from customers, both of which enhance the shopping experience. Such products may either be compared, helping consumers identify the most suitable options, or combined, allowing them to fulfill specific shopping needs. For instance, differences in product quality might prompt customers to choose products with superior composition over less expensive options, while the distinct niches of certain products may lead them to combine various items. When presented with a range of closely related products, customers are either able to select the optimal alternative or find a suitable combination. Thus, related products offer value to customers regardless of whether they serve as substitutes or complements. Consequently, information on product relatedness may be crucial for retailers in guiding customers with limited prior information toward products that best suit their needs. In this context, \cite{Diehl2015} highlight that retailers can aid customers by shelving related products either as substitutes or complements, potentially achieving similarly efficient results.

From a marketing perspective, the relations between products, particularly the distinction between substitutes and complements, are frequently examined through market structure analysis using scanner data (see, e.g., \citealp{Elrod2002}). Recent literature provides various models of market structure explaining relations between a potentially large number of products. \cite{Gabel2019} use a neural network language model to represent products as points in a two-dimensional map, in which the similarity of products is measured by the Euclidean distance. \cite{Ruiz2020} propose a sequential probabilistic model of shopping data and use the symmetrized Kullback–Leibler divergence to identify substitutes. \cite{Chen2020} follow the previous two approaches to identify complements and substitutes based on a low-dimensional space of products. \cite{Tian2021} study product relations in a bipartite product-purchase network and define the substitutability and complementarity measures as cosine similarity. The field of market structure analysis, however, dates back to papers such as \cite{Srivastava1981} and \cite{Lattin1985}.

In addition to these approaches, market basket analysis and network-based methods have long been employed to reveal product relationships through co-purchase patterns. The foundational framework of \cite{Agrawal1993} identifies frequent item sets and conditional purchase dependencies, which later evolved into network-based models where product nodes are linked through purchase co-incidence (see, e.g., \citealp{Kim2012}). Several studies also expand the perspective beyond transaction data. \cite{Netzer2012} use text mining on user-generated data to infer perceived product relations. \cite{Manchanda1999} analyze joint purchase decisions across categories using a multivariate probit model, highlighting interdependence through co-purchases. \cite{Shocker2004} propose a typology of inter-product effects and emphasize the behavioral dimensions of cross-category interactions.

We introduce a novel measure of product relatedness based on shared purchasing patterns. Similarly to \cite{Chen2020} and \cite{Ruiz2020}, our premise is that independent products are unlikely to display similar purchasing patterns, whereas substitutes or complements frequently do. Therefore, shared purchasing patterns among products suggest a substitute or complementary relations. Contrary to \cite{Chen2020} and \cite{Ruiz2020}, our methodology offers a straightforward approach to identify these patterns through analysis of common purchases, which is computationally efficient and easy to interpret, without the need to assume any model of market structure.

Our paper is structured as follows. In Section \ref{sec:corr}, we introduce our measure of product relation correlation and present the results of applying this measure to data from a Czech drugstore retail chain. In Section \ref{sec:xpe}, we explore the relationship between our measure and cross-price elasticity using the same dataset, demonstrating that our measure tends to increase as products become more substitutable or complementary. In Section \ref{sec:cluster}, we examine the application of our product relation correlation measure in product clustering, focusing on hierarchical clustering and identifying the most closely related products within subcategories.

\section{Product Relation Correlation}
\label{sec:corr}

The degree to which products are related is reflected in the similarity of their purchasing patterns, as demonstrated by \cite{Chen2020}. In our approach, we assess the similarity in purchasing patterns between two products by evaluating and comparing their co-purchase connections with other intermediate products. The more alike are co-purchases of intermediate products between two products, the stronger relation we infer about these two products.

\begin{figure}
\begin{center}
\includegraphics[width=\textwidth]{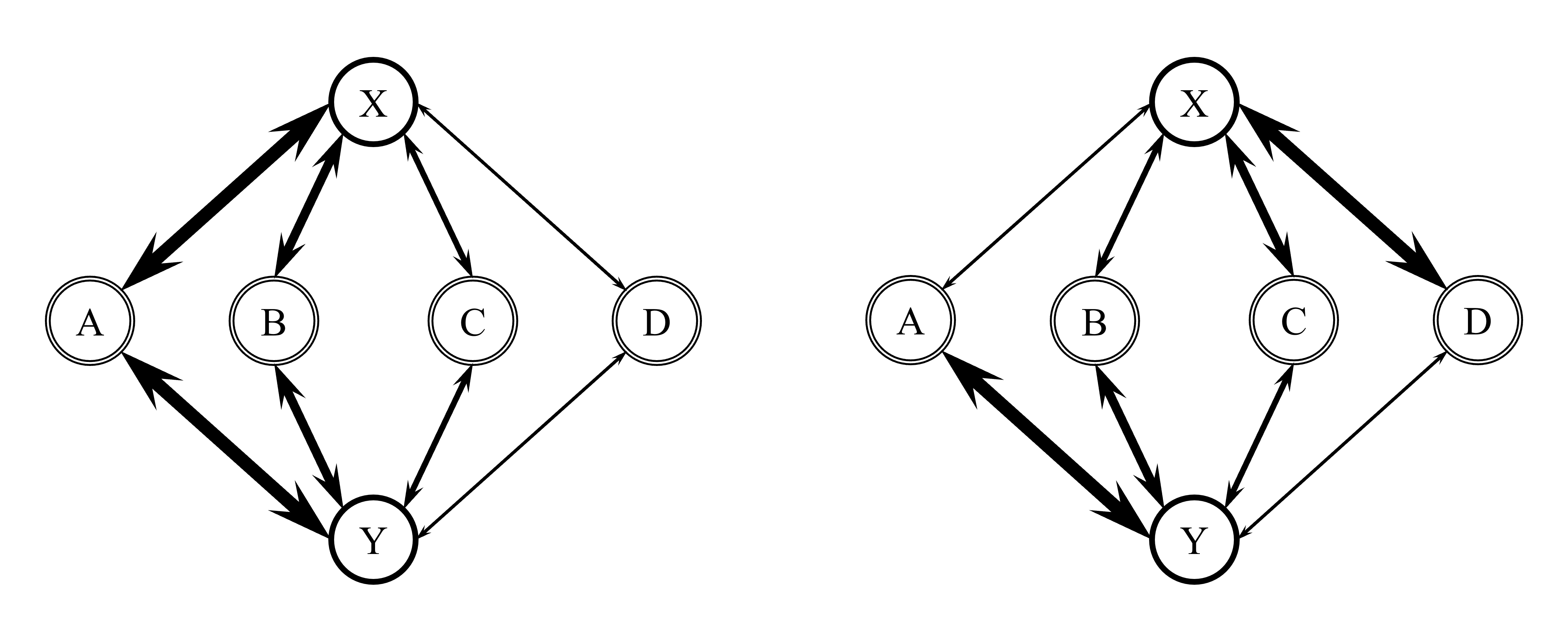}
\caption{Product relation correlation diagram: Two distinct purchasing patterns between products $X$ and $Y$. The strength of the edge represents the co-purchase intensity between product pairs. Left: Strong relation. Right: Weak relation.}
\label{fig:patt}
\end{center}
\end{figure}

Figure \ref{fig:patt} visually represents this concept, using letters to signify products and bold lines to indicate the co-purchase intensity between product pairs. When product X exhibits identical purchases of intermediate products (A, B, C, and D) as product Y, we infer a strong relation between X and Y. This is quantified as a high correlation between their co-purchases of other intermediate products, hence we measure the product relation correlation.

To an extent, shared co-purchases allow for an interpretation that products may share complements.  However, \cite{Ruiz2020} object to this interpretation, arguing that products bought together may be complements or simply purchased based on the consumer’s context, without a necessary complementary relationship. For instance, in their example, diapers and baby formula are often bought together, though they believe these items are not necessarily complementary. We acknowledge this perspective but argue that true complements tend to be co-purchased more frequently than unrelated products, simply due to the nature of complementary items. Therefore, we expect a higher co-purchase intensity between complements, making it plausible to interpret the strongest shared co-purchases as an indication of shared complements.

In this context, products that share complements can either form a complete set of complements or belong to distinct sets of complements. Accordingly, these products may function as complements to each other if they share the same complement set, or as substitutes if they belong to separate sets of complements. For instance, consider items like ham and butter in terms of their complementary products. Baguette and bread both share ham and butter as complements in the context of making a sandwich, yet they typically serve as substitutes because a consumer generally chooses one over the other. In contrast, toast and eggs, in the context of preparing a classic savory breakfast, also share complements like ham and butter but act as complements within the same set. Thus, as highlighted by \cite{Chen2020}, products with shared purchasing patterns may function as either substitutes or complements.

\begin{figure}
\begin{center}
\includegraphics[width=10cm]{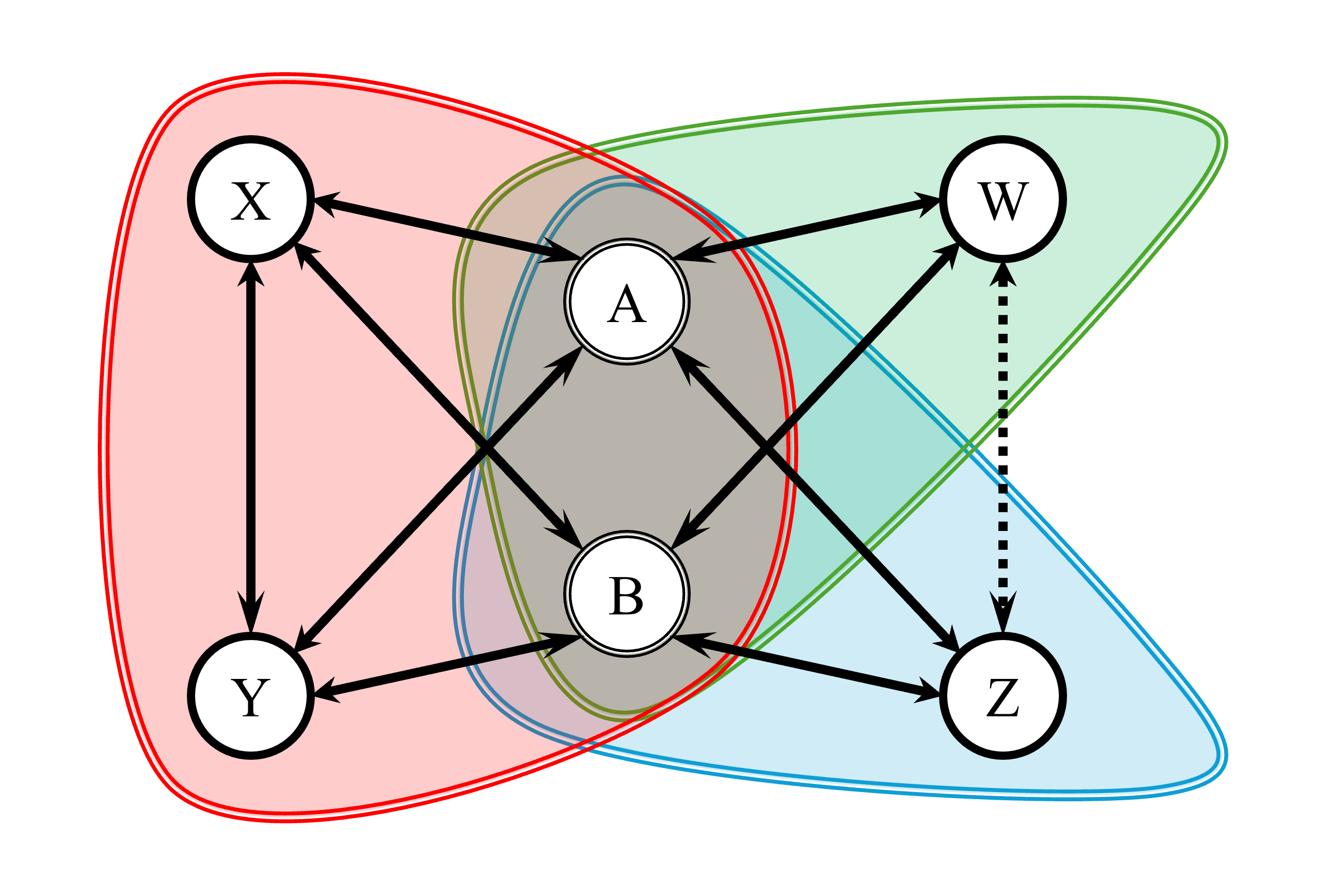}
\caption{Three distinct sets of products with each doubled colored line denoting a distinct group of complements. Illustrating that shared purchase patterns can result in products being either substitutes or complements.}
\label{fig:setpatt}
\end{center}
\end{figure}

This principle is further exemplified in Figure \ref{fig:setpatt}, where each doubled colored line denotes a distinct group of complements. The dashed line between W and Z, representing bread and baguette, denotes that both products are substitutes, whereas full line between X and Y, representing toast and eggs, indicates that they are complements. Products A and B, representing ham and butter, are common complements for the other products.

\subsection{Methodology}

Scanner data contains information on every product sold by a retail store. A set of products bought together by a single customer at a given time is labeled as a shopping basket. The total number of baskets is denoted as $n$, the number of baskets containing product $i$ as $n_i$, $i=1,\ldots,m$, and the number of baskets containing both products $i$ and $j$ as $n_{ij}$, $i,j=1,\ldots,m$. The total number of products is denoted as $m$.

We propose to measure the relation between products $i$ and $j$ by the Pearson correlation coefficient between counts $n_{ik}$ and $n_{jk}$ for $k \neq i,j$. Specifically, we define
\begin{equation}
\label{eq:rho}
\rho_{ij} = \frac{ \sum_{k \neq i,j} \left( n_{ik} - \mu_{i} \right) \left( n_{jk} - \mu_{j} \right) }{ \sqrt{\sum_{k \neq i} \left( n_{ik} - \mu_{i} \right)^2 \sum_{k \neq j} \left( n_{jk} - \mu_{j} \right)^2} } \, , \qquad i,j = 1,\ldots,m \, ,
\end{equation}
where $\mu_{i}$ is given by
\begin{equation}
\mu_{i} = \frac{1}{m - 1} \sum_{k \neq i} n_{ik} \, , \qquad i = 1,\ldots,m \, .
\end{equation}

As $\rho_{ij}$ is a correlation coefficient, it takes values in $[-1,1]$. High values of $\rho_{ij}$ are caused by similar purchase patterns of products $i$ and $j$ with respect to the other products $k$; more specifically, by similar sets of complements to products $i$ and $j$. Similarly to \cite{Chen2020}, we attribute this behavior to either substitutes or complements. Values of $\rho_{ij}$ near zero indicate independent products. High values of $\rho_{ij}$ indicate related products. Negative values of $\rho_{ij}$ indicate products with opposite purchase patterns, suggesting that both products $i$ and $j$ have their own distinct complements. Note that expression \eqref{eq:rho} is identical to the expression when the summations are taken over the whole set of products $k=1, \ldots,m$ with $n_{ii}$ set to $\mu_{i}$. Using this identity, it is then clear that $\rho_{ij}$, $i,j=1,\ldots,m$ form a positive-semidefinite correlation matrix.

Computational simplicity of the proposed approach is obvious. Product relation correlation $\rho_{ij}$ is based only on counts $n$, $n_i$, and $n_{ij}$. When a new basket arrives, it is needed to increase the appropriate counts by one and then recompute equation \eqref{eq:rho}. Besides the resulting correlation, it is required to store $m^2+1$ integers in a database.

Our approach is therefore a streaming algorithm, which can examine a sequence of inputs in a single pass only. This is a great computational advantage as the number of rows in any transaction database increases rapidly. For a survey of the pioneering data stream literature, see \cite{Muthukrishnan2005}. Recent studies in this field include e.g.\ \cite{Cerny2019} who deal with estimation and diagnostics of linear regression and \cite{Holy2023a} who focus on volatility estimation using high-frequency data.

\subsection{Application}

In our empirical analysis, we analyze a Czech drugstore retail chain. The dataset contains $n \sim 20 \text{ million}$ shopping baskets and $m \sim 10,000$ products with at least 500 units sold during the analyzed period. The products are hierarchically divided into $\sim 50$ categories and $\sim 400$ subcategories. We focus on the category \textit{Products for Men}, which contains 12 subcategories and $566$ products. Most drugstore customers are women and the category \textit{Products for Men} is rather a small world of its own.

\begin{figure}
\begin{center}
\includegraphics[height=6cm]{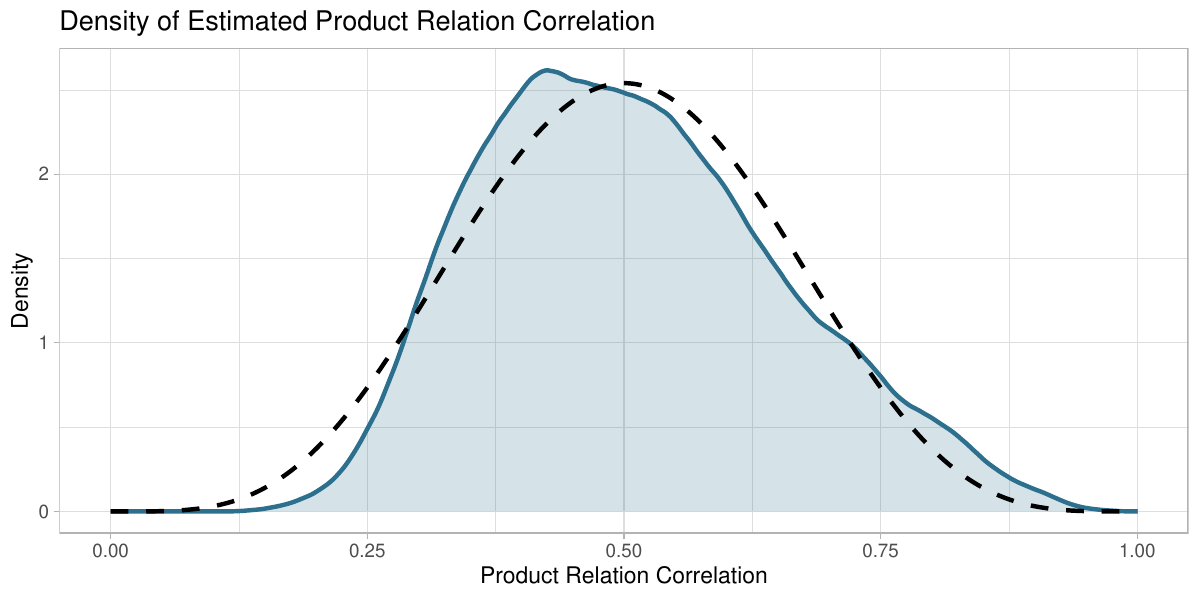}
\caption{The kernel density of $\rho_{ij}$ for the \textit{Products for Men} category with the fitted density of the symmetric beta distribution (dashed).}
\label{fig:density}
\end{center}
\end{figure}

We estimate $\rho_{ij}$ for the pairs of products within the \textit{Products for Men} category using the formula \eqref{eq:rho}. Hence, $i$ and $j$ are indices of products from the \textit{Products for Men} category but we let $k$ take values from the set of all products, not just one specific category.  
In our data, negative values of $\rho_{ij}$ do not occur at all. 
Figure \ref{fig:density} shows the kernel density of the estimated $\rho_{ij}$.  The figure also includes the fitted density of the symmetric beta distribution to visually emphasize the asymmetry of the kernel density.\footnote{First, the kernel density is estimated using the Epanechnikov kernel. Second, the beta distribution is made to be symmetric by setting $\alpha = \beta$ and then parameter $\alpha = \beta = 5.31$ is estimated by the method of moments.} The kernel density of $\rho_{ij}$ is concentrated around mean value 0.51 and is slightly positively skewed. Concerning tails, 0.28 percent of $\rho_{ij}$ is lower than 0.20 and 3.63 percent is higher than 0.80. Densities of $\rho_{ij}$ for other product categories are fairly similar.

\section{Relation to Cross-Price Elasticity of Demand}
\label{sec:xpe}

An essential concept in identifying substitutes and complements is cross-price elasticity of demand, which quantifies the percentage change in demand for one product in response to a percentage change in another product's price, as outlined in foundational microeconomics literature (see, e.g., \citealp{Mankiw2015}). In this straightforward definition, product A is a substitute for product B if an increase in the price of B, ceteris paribus, leads to an increase in the demand for A. Product A is a complement for product B if an increase in the price of B, ceteris paribus, leads to a decrease in the demand for A. Product A is independent of product B if a change in the price of B, ceteris paribus, does not lead to a change in the demand of A. 

To determine the relation between two products and measure its strength, a complex model relating prices, demands, and possibly other variables for multiple products is required. As the number of products increases, so does the dimension of the model causing difficulties in estimation (see, e.g., \citealp{Chernozhukov2019}). Furthermore, prices of products may not be observed for sufficiently long period and detail or may remain constant (see, e.g., \citealp{Ruhm2012}). Finally, there are other factors besides prices which influence the estimation of relation between products such as seasonality and product unavailability (see, e.g., \citealp{Vulcano2012}). For these reasons, other measures of substitutability and complementarity are often adopted, but in our analysis we stick with the cross-price elasticity of demand because of its interpretational and computational simplicity, and established relevance in economic analysis.

Our proposed measure of product relatedness assesses the extent to which a pair of products deviates from independence, indicating they may function as either substitutes or complements. Independent products are assumed to exhibit a cross-price elasticity of zero, whereas substitutes and complements would demonstrate a departure from zero in either a positive or negative direction, respectively.

\subsection{Methodology}

Scanner data includes daily records of quantities purchased for product $i$ by various customers, alongside the prices at which these purchases were made. The aggregated daily quantity of product $i$ purchased across all customers is represented as $\mathbf{x}_i$, while $\mathbf{p}_i$ denotes the average price at which product $i$ was bought each day. The length of $\mathbf{x}_i$ depends on the number of days in which product $i$ was purchased. Thus, more frequently purchased products yield longer vectors, while less common products yield shorter ones. For any two products $i$ and $j$, if $|\mathbf{x}_i| < |\mathbf{x}_j|$, there were days in the dataset where product $j$ was purchased but product $i$ was not. In contrast, price vectors $\mathbf{p}_i$ are all equal in length, reflecting estimated daily average prices, so that $\forall i,k,s \in 1, \dots, m: |\mathbf{x}_i| \leq |\mathbf{p}_k| = |\mathbf{p}_s|$.

Price elasticity of demand, denoted as $\zeta_{ij}$, quantifies the demand response for product $i$ to price changes in product $j$, expressed as the percentage change in quantity demanded given a percentage change in price. This elasticity metric differentiates between own-price elasticity (when $i = j$) and cross-price elasticity (when $i \neq j$). In theory, demand can be viewed as a function of all prices and wealth, i.e., $x_i := x_i(p_1, \dots, p_m, w)$. Consequently, price elasticity is defined as
\begin{equation}
    \zeta_{ij} = \frac{\partial x_i(p_1, \dots, p_m, w)}{\partial p_j} \frac{p_j}{x_i(p_1, \dots, p_m, w)}= \frac{\partial \ln(x_i(p_1, \dots, p_m, w))}{\partial \ln(p_j)} \, ,
    \label{eq:priceel}
\end{equation}
where $\zeta_{ij}$ captures the elasticity of demand for product $i$ with respect to the price of product $j$.

To estimate average demand elasticities, the following simple regression model is employed:
\begin{equation}
    \ln(\mathbf{x}_i) = \alpha_i + \zeta_{ii} \ln(\mathbf{p}_i) + \sum_{j \neq i}^m \zeta_{ij} \ln(\mathbf{p}_j) + \theta \ln(\mathbf{w})  + \mathbf{\epsilon}_i \, , \qquad j = 1,\ldots,m \, ,
    \label{eq:regelas}
\end{equation}
where $\mathbf{w} = \sum_k \mathbf{p}_k \circ \mathbf{x}_k$ represents total daily expenditure, and $\zeta_{ij}$ signifies the cross-price elasticity for product $i$ with respect to product $j$. Positive values of $\zeta_{ij}$ indicate substitutes, while negative values suggest complements.  In general, endogeneity in the regression parameters may be expected due to the mutual relationship between quantity and price (see, e.g., \citealp{Hausman1983}); however, any resulting bias is assumed negligible here given the strong influence suppliers exerted on pricing in this dataset.

In cases with few observations and numerous products, such as our dataset, Ordinary Least Squares (OLS) estimation of the regression model (\ref{eq:regelas}) becomes unsuitable. Instead, we employ the Least Absolute Shrinkage and Selection Operator (LASSO) regression, which is well-suited for such high-dimensional contexts (\citealp{Tibshirani1996}). The LASSO regression applied to our model is formulated as:

\begin{equation}
\label{eq:lasso}
\boldsymbol{\hat{\beta}}_i = \underset{\boldsymbol{\hat{\zeta}}_i, \hat{\alpha}_i, \hat{\theta}}{\arg \min} \Biggl\{ \Biggl( \ln(\mathbf{x}_i) - \hat{\alpha}_{i} - \hat{\zeta}_{ii} \ln(\mathbf{p}_i) - \sum_{j \neq i} \hat{\zeta}_{ij} \ln(\mathbf{p}_j) - \hat{\theta}_i \ln(\mathbf{w}) \Biggl)^2 + \, \lambda \Biggl( \sum_{j=1}^m |\hat{\zeta}_{ij}| + |\hat{\alpha}_{i}| + |\hat{\theta}_i| \Biggl) \Biggl\} \, ,
\end{equation}
where $\boldsymbol{\hat{\beta_i}} = ( \boldsymbol{\hat{\zeta_i}}, \hat{\alpha}_i, \hat{\theta}_i )^T $ is the vector of parameters, containing the constant $\hat{\alpha}_i$, the income elasticity $\hat{\theta}_i$, and the vector of price elasticities $\boldsymbol{\hat{\zeta_i}}= (\hat{\zeta}_{i1}, \hat{\zeta}_{i2}, \dots, \hat{\zeta}_{im})$ for product $i$. 
 
The LASSO regression is estimated repeatedly for each product $i$ in the dataset. The model yields two cross-price elasticity estimates for each product pair: one when $\ln(\mathbf{x}_i)$ is the dependent variable and another when $\ln(\mathbf{x}_j)$ is. Given the analysis focuses on Marshallian demands, symmetry in cross-price elasticities is not guaranteed (see, e.g., \citealp{Irvine1998}), leading to potential discrepancies between $\hat{\zeta}_{ij}$ and $\hat{\zeta}_{ji}$. Consequently, both estimates are considered as independent observations of cross-price elasticities, effectively increasing our sample size. 

The parameter $\lambda$ is critical in LASSO, moderating model complexity and ensuring the elimination of price elasticities for independent products by setting their coefficients to zero. The choice of $\lambda$ directly affects model consistency (\citealp{Tibshirani1996}); hence, its value is determined through cross-validation. First, we cross-validate for each product individually, resulting in a vector $\boldsymbol{\lambda}$ of length $m$, with each $\lambda_i$ corresponding to product $i$. Subsequently, the median of $\boldsymbol{\lambda}$, denoted $\lambda = \text{med}(\boldsymbol{\lambda})$, is used as a universal penalty across all regression models.

\subsection{Application}

The dataset from a Czech drugstore retail chain, specifically focused on the \textit{Products for Men} category, covering one year and including $\approx$ 2 million unique customers. Besides other mentioned variables, the dataset contains information on price at which a product was sold and the amount of purchased quantity.

Two primary issues are identified within this dataset:
\begin{itemize}
    \item[(i)] \emph{Low Variability in Purchased Quantities}: Most transactions involved the purchase of very few product units, with majority of transactions comprising a single unit purchase.
    \item[(ii)] \emph{Personalized Pricing}: The dataset contained personalized prices, accounting for displayed discounts, customer card discounts, or other promotional reductions, complicating the estimation of cross-price elasticity due to the ambiguity in observed prices, as the estimation requires the knowledge of prices on display.
\end{itemize}

\begin{figure}
\begin{center}
\includegraphics[height=6cm]{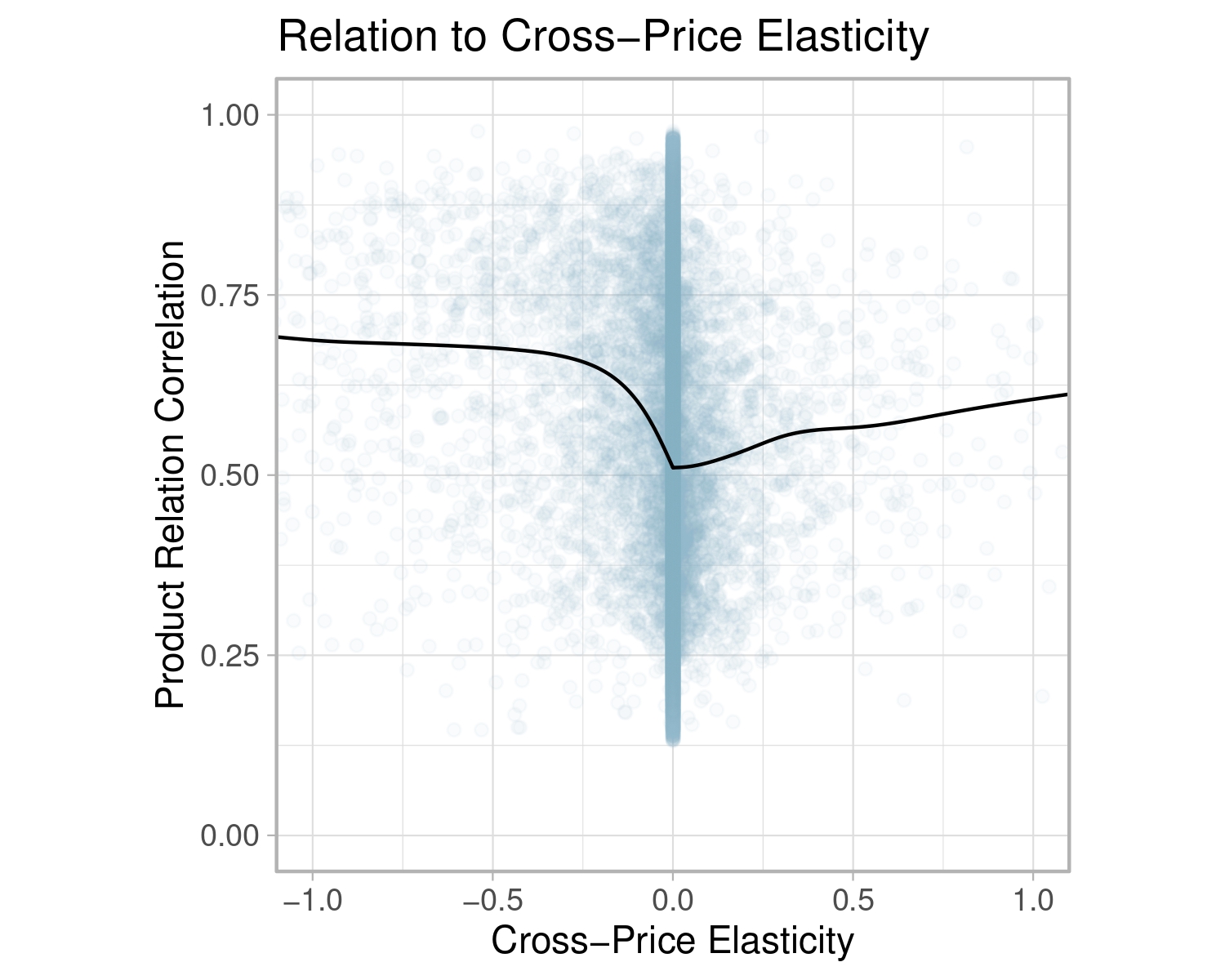}
\caption{Association between product relation correlation and cross-price elasticity: Scatter plot and a smooth spline.}
\label{fig:smooth}
\end{center}
\end{figure}

To address these issues, we aggregated data daily, summing purchased quantities and averaging prices. This approach increased demand variance, minimized accounting discrepancies, and better reflected displayed prices, though it reduced observation counts. Average prices were selected over high quantiles (e.g., 99 \% or 95 \%) for estimating displayed prices, as it better represented the prices encountered by an average consumer, including personalized discounts. We consider this approach to be more reflective of the decision-making process, especially when analyzing aggregated demand.

Using the LASSO regression model (\ref{eq:lasso}), we estimated average cross-price elasticities for all product pairs, including both $\zeta_{ij}$ and $\zeta_{ji}$, and compared them to the product relation correlation $\rho_{ij}$, yielding several hundred thousand paired observations of cross-price elasticity and product relation correlation. The results, visualized in Figure \ref{fig:smooth} as a scatter plot with a fitted smooth spline,\footnote{A cubic smoothing spline with 5 degrees of freedom (the trace of the smoother matrix) is fitted separately for the positive and negative part.} indicate a positive association between product relation correlation and their classification as either substitutes or complements, aligning with findings by \cite{Chen2020}. This implies that products identified as related are more likely to be substitutes or complements.

\begin{figure}
\begin{center}
\includegraphics[width=9cm]{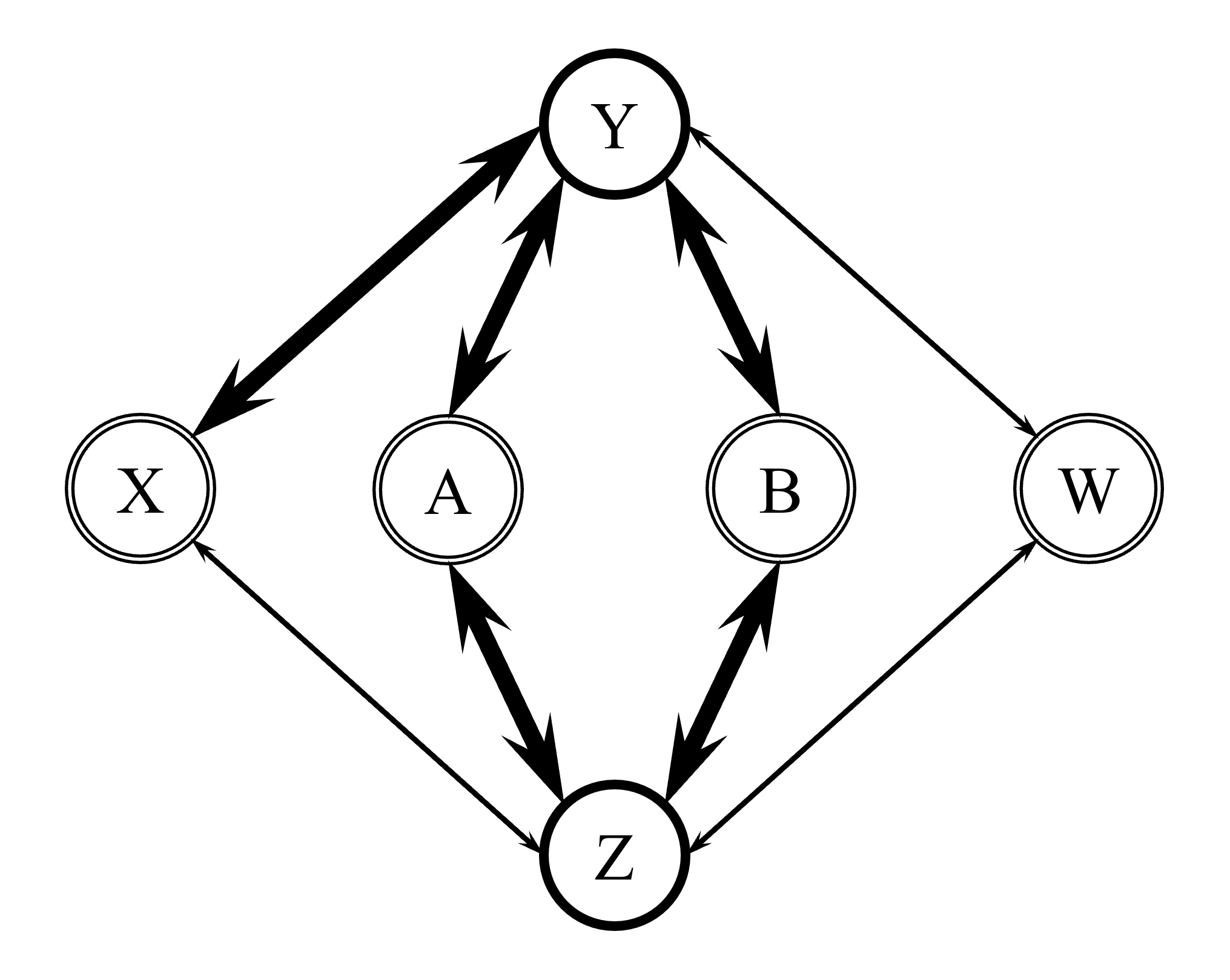}
\caption{Slightly different purchasing patterns between products from two substitutable complement sets of different size. The strength of the edge represents the co-purchase intensity between product pairs.}
\label{fig:caveat}
\end{center}
\end{figure}

The asymmetric association between product relation correlation and cross-price elasticity can be best illustrated using the framework outlined in Figure \ref{fig:setpatt}. Suppose not only W and Z are substitutes, but the entire blue product set serves as a substitute to the entire red product set. In the earlier example, this represents a consumer deciding between a classic savory breakfast and a sandwich, with both sets being substitutes. Our measure compares purchasing patterns between pairs of products rather than whole complement sets. If we measure the product relation correlation between Z and Y, their purchasing patterns may differ slightly because Y is strongly co-purchased with X from the same complement group, unlike Z, as shown in Figure \ref{fig:caveat}.

This limitation in our measure must be considered when estimating the relations between products, particularly in contexts where entire sets of products may serve as substitutes. This issue arises when substitutable relations extend across groups of products rather than between individual items. If substitutable sets differ by only one product, meaning that only single products are substitutable, this limitation does not apply. The consequence is that complements appear to be more related than substitutes.

In cases where sets of different sizes are substitutable, as in our example of classic savory breakfast versus a sandwich, it may even be appropriate that the measured relation between two products of substitutable sets yields lower estimates. This reflects that a single product in one set may act as a substitute for multiple products together in the other set, diluting the perceived substitutability when considered only pairwise.

A possible solution, and a direction for further research, would involve grouping products that are believed to be strongly complementary into a single virtual product. We could then assess the correlation of this virtual product with another product. If the correlation strength between the products increases after grouping, this would suggest that the grouping is justified.

\section{Use in Product Clustering}
\label{sec:cluster}

Clustering of products is a common task in marketing analysis. For example, \cite{Lingras2014} and \cite{Ammar2016} iteratively cluster products and customers based on their interactions. \cite{Holy2017} and \cite{Sokol2021} cluster products using penalty for joint occurrence of products in a shopping basket. Cluster analysis is also used in retail to obtain customer segments or shopping missions (see e.g.\ \citealp{Reutterer2006, Sokol2021a}).

\begin{figure}[t]
\begin{center}
\includegraphics[width=12cm]{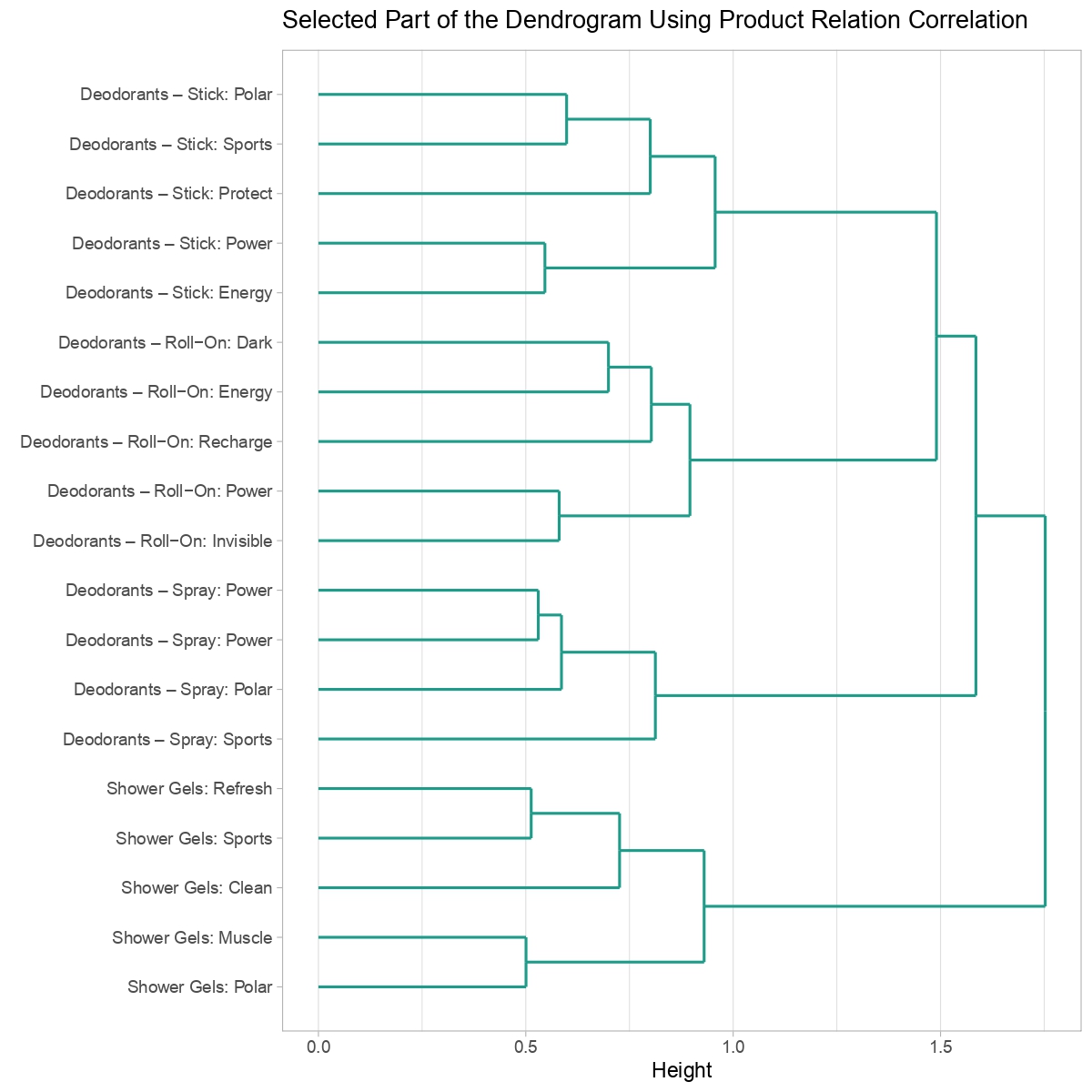}
\caption{The dendrogram of hierarchical clustering for products with the same brand.}
\label{fig:dendro}
\end{center}
\end{figure}

Our measure enables clustering based on product relations, supporting tasks like shelf space optimization. For example, \cite{Diehl2015} demonstrate that grouping products on shelves either by substitutes or complements can be effective, depending on the consumer type entering the store. Thus, product relations are likely a key factor in optimizing the shelf organization.

\subsection{Methodology}
The proposed measure of product relation correlation can be used for clustering of products as well. First, we transform the correlation coefficient $\rho_{ij}$  into a distance using the law of cosines as
\begin{equation}
\label{eq:distance}
d_{ij} = \sqrt{2 - 2\rho_{ij}} \, , \qquad i,j = 1,\ldots,m \, .
\end{equation}
This is a metric as it has the following properties:
\begin{itemize}
\item[(i)] \emph{Identity of Indiscernibles}: $d_{ii}=0$ and $d_{ij}>0$ for $i \neq j$. \footnote{Here, we neglect the possibility of two products always bought together and never apart, for example in some package, as they would be truly indiscernible.}
\item[(ii)] \emph{Symmetry}: $d_{ij}=d_{ji}$.
\item[(iii)] \emph{Triangle Inequality}: $d_{ij} \leq d_{ik} + d_{kj}$.
\end{itemize}
It is also related to the Euclidean distance. Specifically, it is equivalent to the Euclidean distance in the space of standardized vectors. However, it is not equal to the Euclidean distance in the original space unless the vectors are already standardized. Nevertheless, the matrix of distances obtained from \eqref{eq:distance} is Euclidean in the sense that there exists a configuration of points in some Euclidean space having these interpoint distances (see e.g. \citealp{Mardia1979, Gower1986}).

Positively correlated products have small distance while negatively correlated products have high distance. Distance $d_{ij}$ can then be used as a dissimilarity measure in various clustering methods such as k-means and hierarchical clustering (see e.g.\ \citealp{Hastie2008} for a textbook treatment and \citealp{Hennig2022} for a comparative study).

In marketing applications like product listings or discount promotions, the aim of clustering is often to form small groups of related products rather than fewer large groups. A hierarchical structure can be advantageous for the decision-making process, where we can observe, for example, whether a deodorant brand or its scent or other characteristic plays a larger role in the successive clustering. Thus, the behavior of the average customer can be revealed such as whether they prefer a different scent from the same brand or the same (or similar) scent from a different brand. This can be subsequently used when there is a shortage of product for which the customer came.

\subsection{Application}

Applying hierarchical clustering using Ward's criterion, the products are clustered into 566 hierarchical cuts (levels) that form a tree. In particular, we used Ward2 algorithm by \cite{Murtagh2014} with squared distances between products as defined in Equation \eqref{eq:distance} used as a dissimilarity measure. Ward's method aims to minimize the increase in variance when merging clusters, leading to compact and well-separated clusters and it is suitable to use on the distance measure.

The tree cuts correspond to the number of clusters in a given stage. 
For example, at the 30th cut we find a cluster of deodorants of all forms and shower gels with the same brand. Following the cut hierarchy, the shower gels are first separated from the deodorants. Subsequently, deodorants are also separated by form type into spray, stick and roll-on. The individual fragrances are then grouped together in a further subdivision. The described progress is shown in Figure \ref{fig:dendro} with the height representing the distance between clusters at the point they are merged. Ward's method aims to minimize the total within-cluster variance, so the height reflects the increase in variance when two clusters are combined. We have also computed the cophenetic correlation coefficient \citep{Sokal1962} to assess how well the dendrogram preserves the pairwise distances of the original data. The resulting value of 0.834 indicates a strong fit.

\begin{figure}
\begin{center}
\includegraphics[width=\textwidth]{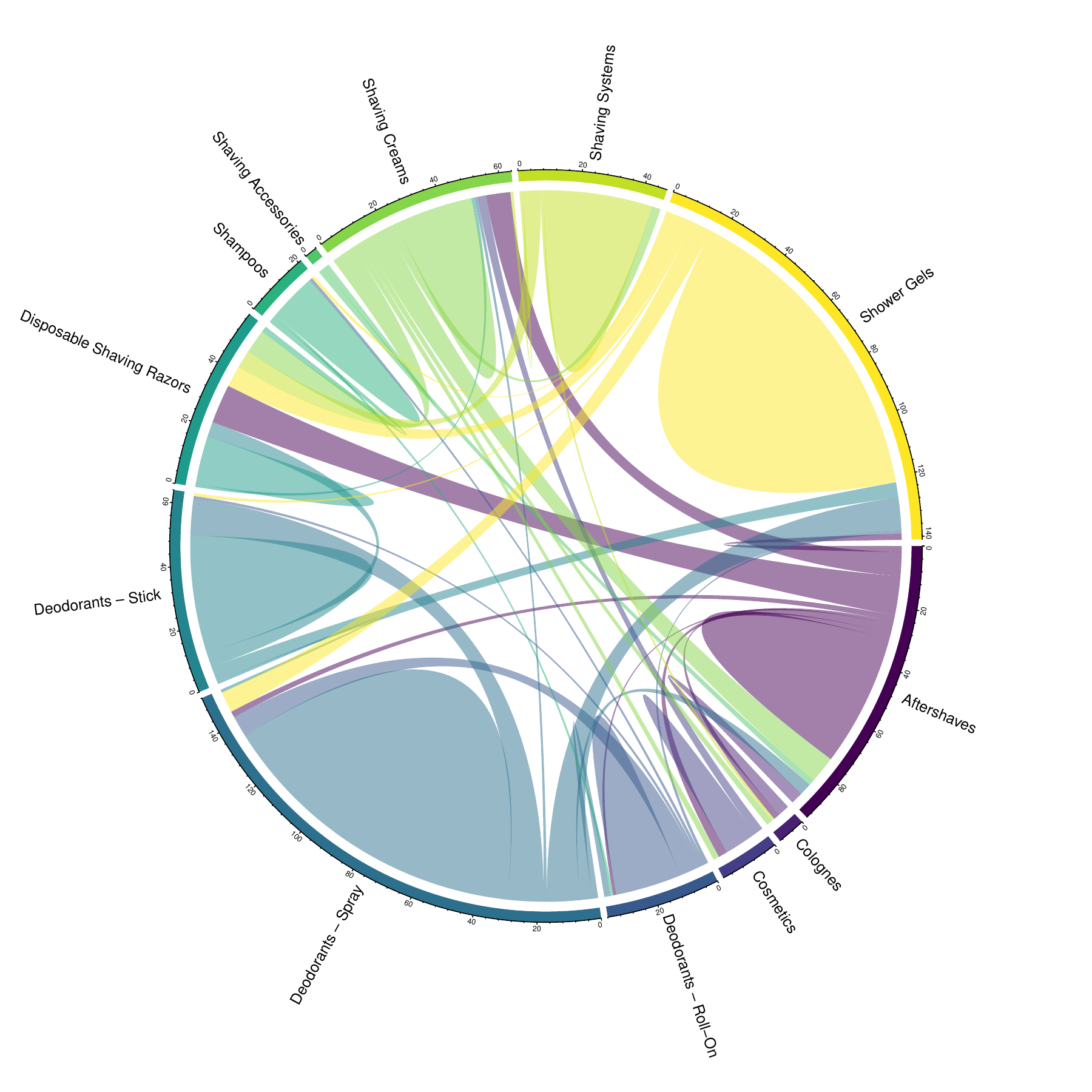}
\caption{The chord diagram of the closest related product by subcategories according to $\rho_{ij}$.}
\label{fig:chord}
\end{center}
\end{figure}

Additionally, we examined the most closely related product for each item (i.e., the product with the highest $\rho_{ij}$), investigating whether the closest matches are within the same subcategory or across different subcategories. The chord diagram in Figure \ref{fig:chord} illustrates these relations, highlighting the most related products within and across subcategories based on the $\rho_{ij}$ measure.  Notably, a significant majority of products (74.73 \%) find their closest relation within the same subcategory. However, there are significant exceptions, such as certain deodorants from the same brand differing only in form (e.g., spray vs. roll-on), or disposable shaving razors and shaving systems under the same brand.

The chord diagram in Figure \ref{fig:chord} further supports the notion that product relation correlation encompasses both substitutes and complements. In general, we would infer that products inside categories usually act as substitutes, while products across categories usually act as complements, with some notable exceptions discussed below. This reasoning assumes that within-category products serve similar purposes, whereas cross-category products typically fulfill distinct shopping needs. For instance, many shower gels serve the same function with minimal functional differences, supporting their substitutability. In contrast, cross-category pairings like disposable razors and aftershaves, or deodorants and shower gels, are often complementary, fulfilling combined purposes. However, some products, though categorized separately, serve similar functions and may act as substitutes, such as stick versus spray deodorants.

These findings contrast somewhat with the association between product relation correlation and cross-price elasticity seen in Figure \ref{fig:smooth}. The majority of the closest related products fall within the same category and thus could potentially be evaluated as substitutes, while only about a quarter of closely related products fall across categories and could be thought of as complements. This contrast appears at odds with the results in Figure \ref{fig:smooth}, which suggests that pairs with higher product relation correlation should, on average, be evaluated more as complements than substitutes due to the asymmetrical nature of the relationship between product relation correlation and cross-price elasticity.

There are two primary explanations for this pattern: First, the complementary cross-category relations are stronger, often ranking among the most related products.  This ties to the observation that substitutes generally exhibit lower values of product relation correlation, as illustrated in Figure \ref{fig:smooth} and \ref{fig:caveat}. Moreover, as shown in Figure \ref{fig:density}, the distribution of product relation correlation is asymmetric and skewed toward lower values. Taken together, these patterns suggest a higher prevalence of substitutes in the dataset. Second, while we considered within-category relations as likely substitutes, this assumption does not hold universally, as there are many exceptions. One common example is within the Shaving Systems category, which includes not only razor handles but also individual replaceable razor blades. These are evaluated as complements in our dataset, a plausible interpretation given that replaceable blades enhance the long-term functionality of razor handles. Another example is found in the Shaving Creams category, which includes both shaving creams or foams as well as pre-shaving creams, which are also evaluated as complements, aligning with intuition. In some cases, even products within the Deodorants category—such as deodorants and antiperspirants—are evaluated as complements. Although this interpretation stretches conventional expectations, it cannot be entirely dismissed, as these products meet slightly different needs, making it conceivable that consumers might use both.

\section{Conclusion}
\label{sec:con}

In this paper, we introduce a new measure, product relation correlation, to assess the extent of relatedness between products based on shared purchasing patterns. This measure aims to identify product pairs that are likely to function as either substitutes or complements. Unlike other methods that rely on complex assumptions about market structure, product relation correlation offers a straightforward and computationally efficient approach, focusing solely on shared co-purchases as an indicator of product relations.

We apply \emph{Product Relation Correlation} to data from a Czech drugstore retail chain, focusing on the \textit{Products for Men} category to illustrate its use. The results illustrate how product relation correlation aligns with the traditional economic measure of cross-price elasticity. By comparing product relation correlation values with cross-price elasticity estimates, we observe a pattern where high values of product relation correlation correspond with pairs clearly identified as substitutes or complements through cross-price elasticity of demand. This finding supports product relation correlation as a useful alternative indicator when traditional means of estimating relations between products (e.g., cross-price elasticity) are not readily available or are difficult to obtain. It is particularly helpful when only a preliminary view of product relations is needed, without requiring specific information on the type of relation (i.e., whether products are substitutes or complements). This level of insight may be sufficient for applications such as shelf space optimization, where understanding general relatedness can guide effective product arrangement.

Furthermore, we apply hierarchical clustering to demonstrate how product relation correlation can be used to organize products effectively. Clustering products based on this measure groups items with similar purchase patterns, providing a natural guide for shelf organization strategies. This approach enables retailers to place related products—whether substitutes or complements—in proximity, potentially enhancing the shopping experience.

Future research could explore the combined use of product relation correlation and a straightforward co-purchase frequency measure, which counts how often two products are bought together. While product relation correlation captures the shared structure of co-purchases between products, this additional measure would provide a direct count of co-purchases for each product pair. This combined approach could improve the ability to distinguish substitutes from complements, enhancing applications in retail analytics and shelf space optimization.

Additionally, an alternative approach to studying consumer choice behavior across broader product categories could involve aggregating known complementary products into single virtual product. By combining two or more products that are assumed to be strong complements into a single virtual product, we could analyze how this changes product relation correlation and use these changes to infer additional insights about the relationships within the grouped products.

\section*{Funding}
\label{sec:fund}

The work on this paper was supported by the Internal Grant Agency of the Prague University of Economics and Business under project F4/27/2020 and the Czech Science Foundation under project 22-19353S.


\begin{thebibliography}{45}
\newcommand{\enquote}[1]{``#1''}
\providecommand{\natexlab}[1]{#1}
\providecommand{\url}[1]{\texttt{#1}}
\providecommand{\urlprefix}{}
\expandafter\ifx\csname urlstyle\endcsname\relax
  \providecommand{\doi}[1]{doi:\discretionary{}{}{}#1}\else
  \providecommand{\doi}{doi:\discretionary{}{}{}\begingroup
  \urlstyle{rm}\Url}\fi
\providecommand{\eprint}[2][]{\url{#2}}

\bibitem[{Agrawal \emph{et~al.}(1993)Agrawal, Imieli{\'n}ski, and
  Swami}]{Agrawal1993}
Agrawal R, Imieli{\'n}ski T, Swami A (1993).
\newblock \enquote{Mining Association Rules Between Sets of Items in Large
  Databases.}
\newblock \emph{ACM SIGMOD Record}, \textbf{22}(2), 207--216.
\newblock ISSN 0163-5808.
\newblock \url{https://doi.org/10.1145/170036.170072}.

\bibitem[{Ammar \emph{et~al.}(2016)Ammar, Elouedi, and Lingras}]{Ammar2016}
Ammar A, Elouedi Z, Lingras P (2016).
\newblock \enquote{Meta-Clustering of Possibilistically Segmented Retail
  Datasets.}
\newblock \emph{Fuzzy Sets and Systems}, \textbf{286}, 173--196.
\newblock ISSN 0165-0114.
\newblock \url{https://doi.org/10.1016/j.fss.2015.07.019}.

\bibitem[{Andreasen(1977)}]{Andreasen1977}
Andreasen AR (1977).
\newblock \enquote{A Taxonomy of Consumer Satisfaction/Dissatisfaction
  Measures.}
\newblock \emph{Journal of Consumer Affairs}, \textbf{11}(2), 11--24.
\newblock ISSN 0022-0078.
\newblock \url{https://doi.org/10.1111/j.1745-6606.1977.tb00612.x}.

\bibitem[{Bakos and Brynjolfsson(1999)}]{Bakos1999}
Bakos Y, Brynjolfsson E (1999).
\newblock \enquote{Bundling Information Goods: Pricing, Profits, and
  Efficiency.}
\newblock \emph{Management Science}, \textbf{45}(12), 1613--1630.
\newblock ISSN 0025-1909.
\newblock \url{https://doi.org/10.1287/mnsc.45.12.1613}.

\bibitem[{{\v C}ern{\'y}(2019)}]{Cerny2019}
{\v C}ern{\'y} M (2019).
\newblock \enquote{Narrow Big Data in a Stream: Computational Limitations and
  Regression.}
\newblock \emph{Information Sciences}, \textbf{486}, 379--392.
\newblock ISSN 0020-0255.
\newblock \url{https://doi.org/10.1016/j.ins.2019.02.052}.

\bibitem[{Chen \emph{et~al.}(2020)Chen, Liu, Proserpio, Troncoso, and
  Xiong}]{Chen2020}
Chen F, Liu X, Proserpio D, Troncoso I, Xiong F (2020).
\newblock \enquote{Studying Product Competition Using Representation Learning.}
\newblock In \emph{Proceedings of the 43rd International ACM SIGIR Conference
  on Research and Development in Information Retrieval},  1261--1268.
  Association for Computing Machinery, New York.
\newblock ISBN 978-1-4503-8016-4.
\newblock \url{https://doi.org/10.1145/3397271.3401041}.

\bibitem[{Chernozhukov \emph{et~al.}(2019)Chernozhukov, Hausman, and
  Newey}]{Chernozhukov2019}
Chernozhukov V, Hausman JA, Newey WK (2019).
\newblock \enquote{Demand Analysis with Many Prices.}
\newblock \url{https://doi.org/10.1920/wp.cem.2019.5919}.

\bibitem[{Diehl \emph{et~al.}(2015)Diehl, {van Herpen}, and
  Lamberton}]{Diehl2015}
Diehl K, {van Herpen} E, Lamberton C (2015).
\newblock \enquote{Organizing Products with Complements versus Substitutes:
  Effects on Store Preferences as a Function of Effort and Assortment
  Perceptions.}
\newblock \emph{Journal of Retailing}, \textbf{91}(1), 1--18.
\newblock ISSN 0022-4359.
\newblock \url{https://doi.org/10.1016/j.jretai.2014.10.003}.

\bibitem[{Dr{\`e}ze \emph{et~al.}(1994)Dr{\`e}ze, Hoch, and Purk}]{Dreze1994}
Dr{\`e}ze X, Hoch SJ, Purk ME (1994).
\newblock \enquote{Shelf Management and Space Elasticity.}
\newblock \emph{Journal of Retailing}, \textbf{70}(4), 301--326.
\newblock ISSN 0022-4359.
\newblock \url{https://doi.org/10.1016/0022-4359(94)90002-7}.

\bibitem[{Elrod \emph{et~al.}(2002)Elrod, Russell, Shocker, Andrews, Bacon,
  Bayus, Carroll, Johnson, Kamakura, Lenk, Mazanec, Rao, and
  Shankar}]{Elrod2002}
Elrod T, Russell GJ, Shocker AD, Andrews RL, Bacon L, Bayus BL, Carroll JD,
  Johnson RM, Kamakura WA, Lenk P, Mazanec JA, Rao VR, Shankar V (2002).
\newblock \enquote{Inferring Market Structure from Customer Response to
  Competing and Complementary Products.}
\newblock \emph{Marketing Letters}, \textbf{13}(3), 221--232.
\newblock ISSN 0923-0645.
\newblock \url{https://doi.org/10.1023/A:1020222821774}.

\bibitem[{Gabel \emph{et~al.}(2019)Gabel, Guhl, and Klapper}]{Gabel2019}
Gabel S, Guhl D, Klapper D (2019).
\newblock \enquote{P2V-MAP: Mapping Market Structures for Large Retail
  Assortments.}
\newblock \emph{Journal of Marketing Research}, \textbf{56}(4), 557--580.
\newblock ISSN 0022-2437.
\newblock \url{https://doi.org/10.1177/0022243719833631}.

\bibitem[{Gower and Legendre(1986)}]{Gower1986}
Gower JC, Legendre P (1986).
\newblock \enquote{Metric and Euclidean Properties of Dissimilarity
  Coefficients.}
\newblock \emph{Journal of Classification}, \textbf{3}(1), 5--48.
\newblock ISSN 0176-4268.
\newblock \url{https://doi.org/10.1007/bf01896809}.

\bibitem[{Grandi \emph{et~al.}(2021)Grandi, Burt, and Cardinali}]{Grandi2021}
Grandi B, Burt S, Cardinali MG (2021).
\newblock \enquote{Encouraging Healthy Choices in the Retail Store Environment:
  Combing Product Information and Shelf Allocation.}
\newblock \emph{Journal of Retailing and Consumer Services}, \textbf{61},
  102522/1--102522/9.
\newblock ISSN 0969-6989.
\newblock \url{https://doi.org/10.1016/j.jretconser.2021.102522}.

\bibitem[{Hastie \emph{et~al.}(2008)Hastie, Tibshirani, and
  Friedman}]{Hastie2008}
Hastie T, Tibshirani R, Friedman J (2008).
\newblock \emph{The Elements of Statistical Learning}.
\newblock Springer, New York.
\newblock ISBN 978-0-387-84857-0.
\newblock \url{https://doi.org/10.1007/978-0-387-84858-7}.

\bibitem[{Hausman(1983)}]{Hausman1983}
Hausman JA (1983).
\newblock \enquote{Specification and Estimation of Simultaneous Equation
  Models.}
\newblock In \emph{Handbook of Econometrics}, Volume~1,  391--448. Elsevier.
\newblock ISBN 978-0-444-86185-6.
\newblock \url{https://doi.org/10.1016/S1573-4412(83)01011-9}.

\bibitem[{Hennig(2022)}]{Hennig2022}
Hennig C (2022).
\newblock \enquote{An Empirical Comparison and Characterisation of Nine Popular
  Clustering Methods.}
\newblock \emph{Advances in Data Analysis and Classification}, \textbf{16},
  201--229.
\newblock ISSN 1862-5347.
\newblock \url{https://doi.org/10.1007/s11634-021-00478-z}.

\bibitem[{Hol{\'y} \emph{et~al.}(2017)Hol{\'y}, Sokol, and {\v
  C}ern{\'y}}]{Holy2017}
Hol{\'y} V, Sokol O, {\v C}ern{\'y} M (2017).
\newblock \enquote{Clustering Retail Products Based on Customer Behaviour.}
\newblock \emph{Applied Soft Computing}, \textbf{60}, 752--762.
\newblock ISSN 1568-4946.
\newblock \url{https://doi.org/10.1016/j.asoc.2017.02.004}.

\bibitem[{Hol{\'y} and Tomanov{\'a}(2023)}]{Holy2023a}
Hol{\'y} V, Tomanov{\'a} P (2023).
\newblock \enquote{Streaming Approach to Quadratic Covariation Estimation Using
  Financial Ultra-High-Frequency Data.}
\newblock \emph{Computational Economics}, \textbf{62}(1), 463--485.
\newblock ISSN 0927-7099.
\newblock \url{https://doi.org/10.1007/s10614-021-10210-w}.

\bibitem[{Hol{\'y} and Tomanov{\'a}(2025)}]{Holy2025e}
Hol{\'y} V, Tomanov{\'a} P (2025).
\newblock \enquote{Estimation of Ornstein-Uhlenbeck Process Using
  Ultra-High-Frequency Data with Application to Intraday Pairs Trading
  Strategy.}
\newblock \emph{Annals of Operations Research}.
\newblock ISSN 0254-5330.
\newblock \url{https://doi.org/10.1007/s10479-025-06855-7}.

\bibitem[{Irvine and Sims(1998)}]{Irvine1998}
Irvine IJ, Sims WA (1998).
\newblock \enquote{Measuring Consumer Surplus with Unknown Hicksian Demands.}
\newblock \emph{American Economic Review}, \textbf{88}(1), 314--322.
\newblock ISSN 0002-8282.

\bibitem[{Kim \emph{et~al.}(2012)Kim, Kim, and Chen}]{Kim2012}
Kim HK, Kim JK, Chen QY (2012).
\newblock \enquote{A Product Network Analysis for Extending the Market Basket
  Analysis.}
\newblock \emph{Expert Systems with Applications}, \textbf{39}(8), 7403--7410.
\newblock ISSN 0957-4174.
\newblock \url{https://doi.org/10.1016/j.eswa.2012.01.066}.

\bibitem[{Kort \emph{et~al.}(2020)Kort, Taboubi, and Zaccour}]{Kort2020}
Kort PM, Taboubi S, Zaccour G (2020).
\newblock \enquote{Pricing Decisions in Marketing Channels in the Presence of
  Optional Contingent Products.}
\newblock \emph{Central European Journal of Operations Research},
  \textbf{28}(1), 167--192.
\newblock ISSN 1435-246X.
\newblock \url{https://doi.org/10.1007/s10100-018-0527-x}.

\bibitem[{Lattin and Mcalister(1985)}]{Lattin1985}
Lattin JM, Mcalister L (1985).
\newblock \enquote{Using a Variety-Seeking Model to Identify Substitute and
  Complementary Relationships Among Competing Products.}
\newblock \emph{Journal of Marketing Research}, \textbf{22}(3), 330--339.
\newblock ISSN 0022-2437.
\newblock \url{https://doi.org/10.1177/002224378502200308}.

\bibitem[{Lee and Ariely(2006)}]{Lee2006}
Lee L, Ariely D (2006).
\newblock \enquote{Shopping Goals, Goal Concreteness, and Conditional
  Promotions.}
\newblock \emph{Journal of Consumer Research}, \textbf{33}(1), 60--70.
\newblock ISSN 0093-5301.
\newblock \url{https://doi.org/10.1086/504136}.

\bibitem[{Leeflang and {Parre{\~n}o-Selva}(2012)}]{Leeflang2012}
Leeflang PSH, {Parre{\~n}o-Selva} J (2012).
\newblock \enquote{Cross-Category Demand Effects of Price Promotions.}
\newblock \emph{Journal of the Academy of Marketing Science}, \textbf{40}(4),
  572--586.
\newblock ISSN 0092-0703.
\newblock \url{https://doi.org/10.1007/s11747-010-0244-z}.

\bibitem[{Lingras \emph{et~al.}(2014)Lingras, Elagamy, Ammar, and
  Elouedi}]{Lingras2014}
Lingras P, Elagamy A, Ammar A, Elouedi Z (2014).
\newblock \enquote{Iterative Meta-Clustering Through Granular Hierarchy of
  Supermarket Customers and Products.}
\newblock \emph{Information Sciences}, \textbf{257}, 14--31.
\newblock ISSN 0020-0255.
\newblock \url{https://doi.org/10.1016/j.ins.2013.09.018}.

\bibitem[{Manchanda \emph{et~al.}(1999)Manchanda, Ansari, and
  Gupta}]{Manchanda1999}
Manchanda P, Ansari A, Gupta S (1999).
\newblock \enquote{The "Shopping Basket": A Model for Multicategory Purchase
  Incidence Decisions.}
\newblock \emph{Marketing Science}, \textbf{18}(2), 95--114.
\newblock ISSN 0732-2399.
\newblock \url{https://doi.org/10.2307/19321195}.

\bibitem[{Mankiw(2015)}]{Mankiw2015}
Mankiw NG (2015).
\newblock \emph{Principles of Microeconomics}.
\newblock Cengage Learning, Stamford.
\newblock ISBN 978-1-285-16590-5.

\bibitem[{Mardia \emph{et~al.}(1979)Mardia, Kent, and Bibby}]{Mardia1979}
Mardia KV, Kent JT, Bibby JM (1979).
\newblock \emph{Multivariate Analysis}.
\newblock Probability and Mathematical Statistics. Academic Press.
\newblock ISBN 978-0-12-471252-2.

\bibitem[{Murtagh and Legendre(2014)}]{Murtagh2014}
Murtagh F, Legendre P (2014).
\newblock \enquote{Ward's Hierarchical Agglomerative Clustering Method: Which
  Algorithms Implement Ward's Criterion?}
\newblock \emph{Journal of Classification}, \textbf{31}(3), 274--295.
\newblock ISSN 0176-4268.
\newblock \url{https://doi.org/10.1007/s00357-014-9161-z}.

\bibitem[{Muthukrishnan(2005)}]{Muthukrishnan2005}
Muthukrishnan S (2005).
\newblock \enquote{Data Streams: Algorithms and Applications.}
\newblock \emph{Foundations and Trends in Theoretical Computer Science},
  \textbf{1}(2), 117--236.
\newblock ISSN 1551-3114.
\newblock \url{https://doi.org/10.1561/0400000002}.

\bibitem[{Nelson(1970)}]{Nelson1970}
Nelson P (1970).
\newblock \enquote{Information and Consumer Behavior.}
\newblock \emph{Journal of Political Economy}, \textbf{78}(2), 311--329.
\newblock ISSN 0022-3808.
\newblock \url{https://doi.org/10.1086/259630}.

\bibitem[{Nelson(1974)}]{Nelson1974}
Nelson P (1974).
\newblock \enquote{Advertising as Information.}
\newblock \emph{Journal of Political Economy}, \textbf{82}(4), 729--754.
\newblock ISSN 0022-3808.
\newblock \url{https://doi.org/10.1086/260231}.

\bibitem[{Netzer \emph{et~al.}(2012)Netzer, Feldman, Goldenberg, and
  Fresko}]{Netzer2012}
Netzer O, Feldman R, Goldenberg J, Fresko M (2012).
\newblock \enquote{Mine Your Own Business: Market-Structure Surveillance
  Through Text Mining.}
\newblock \emph{Marketing Science}, \textbf{31}(3), 521--543.
\newblock ISSN 0732-2399.
\newblock \url{https://doi.org/10.1287/mksc.1120.0713}.

\bibitem[{Reutterer \emph{et~al.}(2006)Reutterer, Mild, Natter, and
  Taudes}]{Reutterer2006}
Reutterer T, Mild A, Natter M, Taudes A (2006).
\newblock \enquote{A Dynamic Segmentation Approach for Targeting and
  Customizing Direct Marketing Campaigns.}
\newblock \emph{Journal of Interactive Marketing}, \textbf{20}(3-4), 43--57.
\newblock ISSN 1094-9968.
\newblock \url{https://doi.org/10.1002/dir.20066}.

\bibitem[{Ruhm \emph{et~al.}(2012)Ruhm, Jones, McGeary, Kerr, Terza,
  Greenfield, and Pandian}]{Ruhm2012}
Ruhm CJ, Jones AS, McGeary KA, Kerr WC, Terza JV, Greenfield TK, Pandian RS
  (2012).
\newblock \enquote{What U.S. Data Should Be Used to Measure the Price
  Elasticity of Demand for Alcohol?}
\newblock \emph{Journal of Health Economics}, \textbf{31}(6), 851--862.
\newblock ISSN 0167-6296.
\newblock \url{https://doi.org/10.1016/j.jhealeco.2012.08.002}.

\bibitem[{Ruiz \emph{et~al.}(2020)Ruiz, Athey, and Blei}]{Ruiz2020}
Ruiz FJR, Athey S, Blei DM (2020).
\newblock \enquote{SHOPPER: A Probabilistic Model of Consumer Choice with
  Substitutes and Complements.}
\newblock \emph{Annals of Applied Statistics}, \textbf{14}(1), 1--27.
\newblock ISSN 1932-6157.
\newblock \url{https://doi.org/10.1214/19-aoas1265}.

\bibitem[{Shocker \emph{et~al.}(2004)Shocker, Bayus, and Kim}]{Shocker2004}
Shocker AD, Bayus BL, Kim N (2004).
\newblock \enquote{Product Complements and Substitutes in the Real World: The
  Relevance of ``Other Products''.}
\newblock \emph{Journal of Marketing}, \textbf{68}(1), 28--40.
\newblock ISSN 0022-2429.
\newblock \url{https://doi.org/10.1509/jmkg.68.1.28.24032}.

\bibitem[{Sokal and Rohlf(1962)}]{Sokal1962}
Sokal RR, Rohlf FJ (1962).
\newblock \enquote{The Comparison of Dendrograms by Objective Methods.}
\newblock \emph{Taxon}, \textbf{11}(2), 33--40.
\newblock ISSN 0040-0262.
\newblock \url{https://doi.org/10.2307/1217208}.

\bibitem[{Sokol and Hol{\'y}(2021{\natexlab{a}})}]{Sokol2021}
Sokol O, Hol{\'y} V (2021{\natexlab{a}}).
\newblock \enquote{Clustering with Penalty for Joint Occurrence of Objects:
  Computational Aspects.}
\newblock \url{https://arxiv.org/abs/2102.01424}.

\bibitem[{Sokol and Hol{\'y}(2021{\natexlab{b}})}]{Sokol2021a}
Sokol O, Hol{\'y} V (2021{\natexlab{b}}).
\newblock \enquote{The Role of Shopping Mission in Retail Customer
  Segmentation.}
\newblock \emph{International Journal of Market Research}, \textbf{63}(4),
  454--470.
\newblock ISSN 1470-7853.
\newblock \url{https://doi.org/10.1177/1470785320921011}.

\bibitem[{Srivastava \emph{et~al.}(1981)Srivastava, Leone, and
  Shocker}]{Srivastava1981}
Srivastava RK, Leone RP, Shocker AD (1981).
\newblock \enquote{Market Structure Analysis: Hierarchical Clustering of
  Products Based on Substitution-in-Use.}
\newblock \emph{Journal of Marketing}, \textbf{45}(3), 38--48.
\newblock ISSN 0022-2429.
\newblock \url{https://doi.org/10.2307/1251540}.

\bibitem[{Tian \emph{et~al.}(2021)Tian, Lautz, Wallis, and
  Lambiotte}]{Tian2021}
Tian Y, Lautz S, Wallis AOG, Lambiotte R (2021).
\newblock \enquote{Extracting Complements and Substitutes from Sales Data: A
  Network Perspective.}
\newblock \emph{EPJ Data Science}, \textbf{10}, 45/1--45/27.
\newblock ISSN 21931127.
\newblock \url{https://doi.org/10.1140/epjds/s13688-021-00297-4}.

\bibitem[{Tibshirani(1996)}]{Tibshirani1996}
Tibshirani R (1996).
\newblock \enquote{Regression Shrinkage and Selection via the Lasso.}
\newblock \emph{Journal of the Royal Statistical Society: Series B
  (Methodological)}, \textbf{58}(1), 267--288.
\newblock ISSN 0035-9246.
\newblock \url{https://doi.org/10.2307/2346178}.

\bibitem[{Vulcano \emph{et~al.}(2012)Vulcano, Van~Ryzin, and
  Ratliff}]{Vulcano2012}
Vulcano G, Van~Ryzin G, Ratliff R (2012).
\newblock \enquote{Estimating Primary Demand for Substitutable Products from
  Sales Transaction Data.}
\newblock \emph{Operations Research}, \textbf{60}(2), 313--334.
\newblock ISSN 0030-364X.
\newblock \url{https://doi.org/10.1287/opre.1110.1012}.

\end{thebibliography}

\end{document}